\documentclass[aps,pre,reprint,superscriptaddress,10pt]{revtex4-2}
\usepackage{amsmath}
\usepackage{graphicx}% Include figure files
\usepackage{dcolumn}% Align table columns on decimal point
\usepackage{bm}% bold math
\usepackage{hyperref}
\usepackage{courier}
\usepackage{braket}
\usepackage{float}
\usepackage[usenames,dvipsnames]{xcolor}
\hypersetup{
  colorlinks   = true,	%Colours links instead of ugly boxes
  urlcolor     = blue,	%Colour for external hyperlinks
  linkcolor    = blue,	%Colour of internal links
  citecolor   = blue	%Colour of citations#NavyBlue
}
\begin{document}
\newcommand{\ve}{\varepsilon}
\title{Quantum manifestations of homogeneous and inhomogeneous  oscillation suppression states}

\author{Biswabibek Bandyopadhyay}
\affiliation{Chaos and Complex Systems Research Laboratory, Department of Physics, University of Burdwan, Burdwan 713
  104, West Bengal, India}
\author{Taniya Khatun}
\affiliation{Chaos and Complex Systems Research Laboratory, Department of Physics, University of Burdwan, Burdwan 713
  104, West Bengal, India}
\author{Debabrata Biswas}
\affiliation{Department of Physics, Bankura University, Bankura 722 155, West Bengal, India}
\author{Tanmoy Banerjee}
\email[]{tbanerjee@phys.buruniv.ac.in}
%\thanks{}
\affiliation{Chaos and Complex Systems Research Laboratory, Department of Physics, University of Burdwan, Burdwan 713
  104, West Bengal, India}

\date{\today}

\begin{abstract}
We study the quantum manifestations of homogeneous and inhomogeneous  oscillation suppression states in  coupled {\it identical} quantum oscillators. We consider quantum van der Pol oscillators coupled via weighted mean-field diffusive coupling and using the formalism of open quantum system we show that depending upon the coupling and the density of mean-field, two types of quantum amplitude death occurs, namely squeezed and nonsqueezed quantum amplitude death. Surprisingly, we find that the inhomogeneous  oscillation suppression state (or the oscillation death state) does not occur in the quantum oscillators in the classical limit. 
However, in the deep quantum regime we discover an oscillation death-like state which is manifested in the phase space through the symmetry-breaking bifurcation of Wigner function. Our results also hint towards the possibility of the transition from quantum amplitude death to oscillation death state through the ``quantum" Turing-type bifurcation. We believe that the observation of quantum oscillation death state will deepen our knowledge of symmetry-breaking dynamics in the quantum domain. 
\end{abstract}

\maketitle

\section{Introduction}
The collective dynamics of coupled oscillators are of great interest in the field of physics, chemistry, and biology \cite{sync}. The two most important emergent behaviors shown by a system of coupled oscillators are synchronization \cite{piko} and oscillation quenching \cite{kosprep}. 
While synchronization is predominantly governed by the phase dynamics, the oscillation quenching is a manifestation of the amplitude dynamics.  
%Synchronization, which is predominantly governed by the phase dynamics, is observed in a plethora of systems ranging from natural entities to man made devices \cite{piko,grid1}.  

In recent years synchronization in quantum regime has attracted much attention: two seminal papers by \citet{lee_prl} and \citet{brud_prl1} unravel the important aspects of quantum synchronization (i.e., the manifestation of synchronization in the quantum regime) using the paradigmatic quantum van der Pol oscillators. Later on, several studies explored the richness of quantum synchronization \cite{lee_pre_1,brud-ann15,brudprl_17,morgan,schoell_qm} and proposed techniques to improve synchronization measures in the background of quantum noise \cite{squeezing,enhance-kwek}. Recent experimental observations of synchronization in the quantum regime in spin-1 limit-cycle oscillators \cite{expt1} and IBM-Q system \cite{expt2} established that quantum synchronization is a physical reality.

Although much attention has been given in revealing the quantum manifestation of synchronization, the oscillation quenching states is relatively a less explored topic. \citet{qad1} first explored the notion of ``quantum" amplitude death in quantum van der Pol oscillators under diffusive coupling. In the classical sense, in the amplitude death (AD) state oscillators arrive at a common steady state which was unstable in the absence of coupling, therefore, AD leads to a stable homogeneous steady state (HSS). Unlike classical AD, the authors showed that the presence of quantum noise hinders the genuine AD state, however, a sufficient decrease in the mean phonon number was considered as the indication of the quantum AD state. Later, \citet{qad2} reported quantum AD in the presence of Kerr type nonlinearity and showed that anharmonicity leads to true quantum effects in the oscillation suppression phenomenon. In Ref.~\cite{qad1} parameter mismatch was introduced explicitly to induce AD and in Ref.~\cite{qad2} the presence of Kerr type nonlinearity effectively introduces frequency detuning between the oscillators that leads to noise induced quantum AD. Therefore, parameter mismatch seems to be a necessary ingredient to induce quantum AD.
%: quantum AD in {\it identical} coupled quantum oscillators has not been studied yet. 

Moreover, in the context of coupled oscillators, oscillation quenching process is much more subtle. Apart from AD there exists another oscillation quenching process, namely  oscillation death (OD) \cite{kosprep}. In the OD state, oscillators populate different coupling dependent nontrivial steady states and thereby give rise to {\it symmetry-breaking} stable inhomogeneous steady states (IHSS). In this context, \citet{kosprl} established that AD and OD may occur in the same system and AD transforms into OD through a symmetry-breaking bifurcation, which resembles the Turing-type bifurcation of spatially extended system \cite{turing}.
However, the quantum mechanical analog of the OD state has hitherto not been reported.

Motivated by the above discussion, in this paper we ask the following questions: (i) What are the different manifestations of quantum amplitude death state in coupled {\it identical} oscillators? (ii) Does OD occur in quantum oscillators? If yes, what is the quantum mechanical analog of an OD state? 
%(ii) What is the quantum mechanical manifestation of the classical Turing-type transition from AD to OD?   
To answer these questions we consider two quantum van der Pol (vdP) oscillators \cite{vdp} coupled by weighted mean-field coupling. The paradigmatic quantum vdP oscillator has been chosen as the test bed to study several emergent dynamics in the quantum domain. More importantly, the quantum vdP oscillators are proposed to be realizable in experiment with trapped ion and ``membrane-in-the-middle" set up \cite{lee_prl,brud_prl1}. The choice of weighted {\it mean-field diffusive} coupling as the coupling scheme adopted in this study is motivated by the fact that it is the simplest yet physically relevant model to distinctly observe AD and OD \cite{tanpre1,tanpre2}. Under normal {\it diffusive} coupling AD appears under parameter mismatch \cite{asen,prasad1}, and OD generally coexists with limit cycle(s) making AD impossible and OD difficult to observe in {\it identical} oscillators \cite{kosprep}. In the present paper, we use two types of weighted mean-field diffusive coupling, namely nonscalar and scalar coupling \cite{scalar}. The nonscalar coupling is known to induce AD only (no OD state is possible) and the scalar coupling is conducive to both AD and OD \cite{scalar-kurths}. 

At this point it is important to understand the difficulty of identifying OD in quantum systems. In the case of AD, the oscillators populate the {\it zero} steady state, therefore, a pronounced reduction in the mean phonon number or increased probability of ground Fock level are the sufficient  indicators of transition from oscillation to quantum AD state \cite{qad1,qad2}. However, in the case of classical OD since two or more than two {\it non-zero} steady states are created, therefore, the mean phonon number and Fock level distribution can no longer distinguish quantum OD and oscillatory states unambiguously.
%are no longer act as the indicative quantifiers of the quantum OD state as they fail to distinguish OD and oscillatory states unambiguously.
Because, in the quantum OD state the mean phonon number does not reduce drastically and the ground state is no longer the highest populated state. 
Therefore, we have to rely largely on the phase space representation: for the limiting case of two oscillators, in the classical OD state two steady states are created, which are {\it displaced} from the origin in phase space. Therefore, in the quantum OD state it is instructive to observe the equivalent {\it displacement} in the Wigner distribution function in the phase space.

In this paper, using the formalism of open quantum systems, we show that under the weighted mean-field diffusive coupling, {\it identical} quantum vdP oscillators exhibit quantum amplitude death. We identify two types of quantum AD states, namely squeezed and non-squeezed quantum AD: the former AD state has not been observed in the previous studies \cite{qad1,qad2}. The quantum AD state is explored using quantum master equation and compared with the AD state of the classical and semiclassical cases. 
%Similar to the parameter mismatch case of \cite{qad1} we find that in the quantum AD state average amplitude is less than that of the noisy classical case. 
Further, we find that the quantum OD state does not occur in quantum oscillators in the {\it classical limit}. However, in the deep quantum region we discover an oscillation death-like state which emerges as the result of the symmetry-breaking bifurcation of Wigner distribution function. Also, we see that the transition from quantum AD to OD provides a qualitative indication of the quantum mechanical analog of the Turing-type bifurcation. 

The rest of the paper is organized in the following manner. The next section describes the classical and quantum van der Pol oscillator. In Sec.~\ref{classical} we describe the mathematical model of classical vdP oscillators coupled through weighted mean-field diffusive coupling. For a clear understanding of the classical dynamics we revisit the bifurcation scenarios that lead to classical amplitude and oscillation death. Section.~\ref{sec:qad} presents the results of quantum amplitude death under nonscalar coupling; also, we compare the results with the noisy classical model. Section~\ref{sec:qod} reports the appearance of squeezed quantum AD and the quantum manifestation of the oscillation death state that appears under scalar coupling. Finally, we conclude the paper in Sec.~\ref{sec:con} discussing the importance of the results.

%\begin{figure}
%%\vspace{0.4cm}
%\includegraphics[width=.48\textwidth]{phase_space}
%\caption{ Quantum limit cycle from quantum van der Pol oscillator. $\omega=2$, $k_1=1$, and $k_2=0.2$.}
%\label{phase_space}
%\end{figure}

\section{van der Pol oscillator: Classical and quantum}
\label{sec:vdp}
%==========================(xy equation)==============================
A van der Pol oscillator has the following mathematical form \cite{vdp}:
\begin{equation}
\label{vdp}
\ddot{x}=-\omega^2x+k_1\dot{x}-8k_2x^2\dot{x},
\end{equation}
where $\omega$ is the intrinsic frequency and $k_1$ is the gain rate corresponding to the linear pumping and $k_2$ is the loss rate corresponding to the nonlinear damping ($k_1,k_2>0$).
%==========================(Amplitude equation)==============================
We can express Eq.~\eqref{vdp} in terms of a complex amplitude $\alpha=x+iy$ (where $\dot{x} = \omega y$) and the corresponding amplitude equation is given by (see Appendix \ref{app:A}):
\begin{equation}
\label{single_amp}
\dot{\alpha}=-i\omega\alpha+(\frac{k_1}{2}-k_2|\alpha|^2)\alpha.
\end{equation}
The oscillator shows a limit cycle oscillation with an amplitude $\sqrt{\frac{k_1}{2k_2}}$.

%==========================(Master equation)==============================
The quantum van der Pol oscillator is represented by the quantum master equation in density matrix $\rho$ \cite{lee_prl,brud_prl1}:
\begin{equation}
\label{single_master}
\dot{\rho}=-i[\omega a^\dag a,\rho]+k_1\mathcal{D}[a^\dag](\rho)+k_2\mathcal{D}[a^2](\rho),
\end{equation}
where $\mathcal{D}[\hat{L}](\rho)$ is the Lindblad dissipator having the form $\mathcal{D}[\hat{L}](\rho)=\hat{L}\rho\hat{L}^\dag-\frac{1}{2}\{\hat{L}^\dag \hat{L},\rho\}$, where $\hat{L}$ represents an operator.  Here and throughout the paper we take $\hbar=1$. $a$ and $a^\dag$ are the Bosonic anihilation and creation operators, respectively. $k_1$ and $k_2$ have the same meaning as the classical case. In the classical limit, linear pumping dominates over the nonlinear damping (i.e., $k_1>k_2$) and one approximates $\langle a\rangle \equiv \alpha$, and starting from the master equation \eqref{single_master} one arrives at the classical amplitude equation \eqref{single_amp} by the following relation: $\dot{\braket{a}}=\mbox{Tr}(\dot \rho a)$ (see Appendix \ref{app:B}).
%Figure~\ref{phase_space} shows the phase space representation of the quantum limit cycle.

%####################### (Section) ############################
\section{Classical VdP oscillators: nonscalar and scalar coupling}\label{classical}
We consider two identical classical van der Pol oscillators, which are coupled via weighted mean-field diffusive coupling scheme. The mathematical model is given below,
%~~~~~~~~~~~~~~~~~~~~~~~~~~~~~~~~~~~~~~~~~~~~b_eqn
\begin{subequations}
\label{classical_eqn}
\begin{align}
\dot{x}_j&=\omega y_j+\varepsilon_1\left(\frac{q}{2}\sum_{m=1}^2x_m-x_j\right), \\
%\end{align}
%\begin{align}
\dot{y}_j&=-\omega x_j+(k_1-8k_2 {x_j}^2)y_j+\varepsilon_2\left(\frac{q}{2}\sum_{m=1}^2y_m-y_j\right),
\end{align}
\end{subequations}
%~~~~~~~~~~~~~~~~~~~~~~~~~~~~~~~~~~~~~~~~~~~~e_eqn
$j\in\{1, 2\}$. $\varepsilon_{1,2}$ are the coupling parameters ($\varepsilon_{1,2}>0$). Both the oscillators have the common eigen-frequency $\omega$.
The control parameter $q$ determines the density of the weighted mean-field. Originally, the parameter $q$ was introduced in the context of quorum sensing in genetic oscillators that controls the extracellular autoinducer concentration in cell to cell communication \cite{qstr,qstr2}. Later on its effect was investigated in physical systems \cite{tanpre1,tanpre2,tanryth} and ecological network  (as an parameter controlling additional mortality) \cite{bandutta}. The coupling scheme was also realized experimentally in electronic circuits \cite{tanpre2,tanryth}. From physical point of view $q$ determines the degree of dissipation in the coupling path: lesser $q$ implies greater dissipation and vice versa. Generally, $q$ acts as a {\it dilution} parameter in the limit  $0\le q\le1$. However, this limit on $q$ is not strict \cite{volkovq}: for $q>1$, it acts as an {\it amplification} parameter. 
%In this paper we will show that $q>1$ gives rise to several novel quantum dynamics in the deep quantum regime.

The coupling scheme of Eq.~\eqref{classical_eqn} can be categorized into two types. (i) {\bf Nonscalar coupling}: When $\varepsilon_{1,2} \ne 0$ the coupling is said to be nonscalar coupling. This type of coupling is conducive for the amplitude death state \cite{scalar-kurths}. (ii) {\bf Scalar coupling}: If either $\ve_{1}=0$ or $\ve_2=0$ the coupling is said to be scalar coupling. In this paper we consider $\ve_1=\ve \ne 0$ and $\ve_2=0$ as this type of ``real part coupling" is known to induce oscillation death \cite{scalar-kurths,scholl-epl}.

\subsection{Nonscalar coupling: Classical AD}
\label{nonscalarad}
The amplitude equation of Eq.~\eqref{classical_eqn} under nonscalar coupling is given by,
\begin{equation}
\label{amp_eqn-vec}
\begin{split}
\dot{\alpha_j}&=-i\omega\alpha_j+(\frac{k_1}{2}-k_2|\alpha_j|^2)\alpha_j+\varepsilon\left(\frac{q}{2}\sum_{m=1}^2\alpha_m-\alpha_j\right).
\end{split}
\end{equation}
Here without any loss of generality we consider $\ve_1=\ve_2=\ve$. 
The system represented in equation \eqref{amp_eqn-vec} has a trivial fixed point at the origin: $\mathcal{F}_{HSS} \equiv (0,0,0,0)$. 
One can evaluate the (inverse) Hopf bifurcation point through which amplitude death appears: 
$\varepsilon_{\mbox{HB,ns}}=\frac{k_1}{2(1-q)}$.

\begin{figure}
%\vspace{0.4cm}
\includegraphics[width=.49\textwidth]{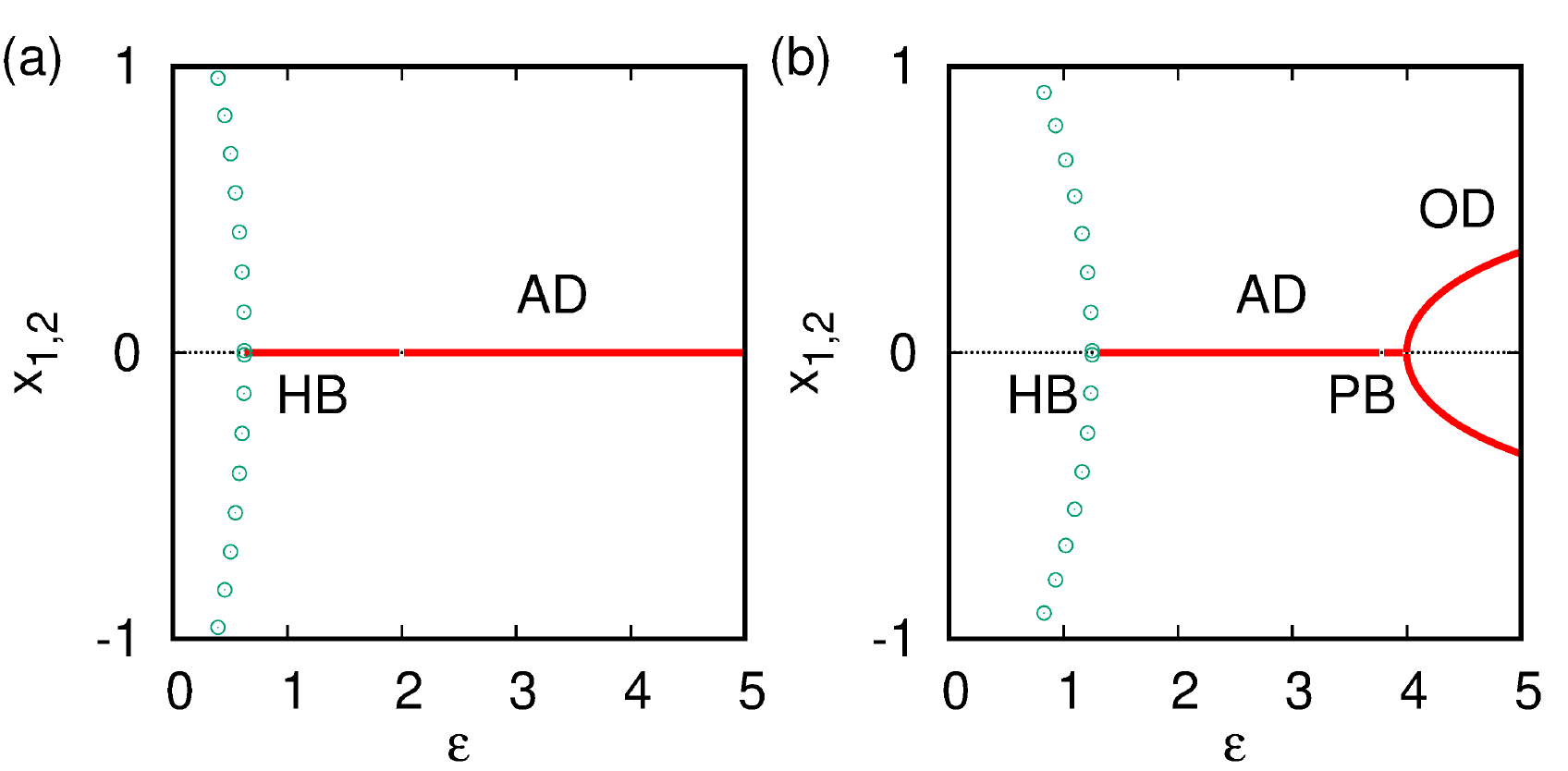}
\caption{ (a) Nonscalar coupling: Amplitude death (AD) occurs through inverse Hopf bifurcation (HB). (b) Scalar coupling: AD appears through HB and oscillation death (OD) emerges through a symmetry-breaking pitchfork bifurcation (PB). Solid (red) line: stable steady state, dotted (black) line: unstable steady state, and hollow circles (green): stable limit cycle. Parameters are $q=0.2$, $\omega=2$, $k_1=1$, and $k_2=0.2$.}
\label{classical_bifur}
\end{figure}

Figure~\ref{classical_bifur}(a) illustrates this scenario in bifurcation diagram of $x_{1,2}$ with $\varepsilon$ for an exemplary parameter set ($\omega=2$, $k_1=1$, $k_2=0.2$, and $q=0.2$) (using \texttt{XPPAUT} \cite{xpp}). In the AD state one has $x_{1,2}=y_{1,2}=0$, i.e., both the oscillators attain a common steady state $\mathcal{F}_{HSS}$ which is the origin.
%The averaged amplitude of oscillation (averaged over 1000 real time) is plotted with increasing coupling parameter. The amplitude of the oscillation shrinks to zero above the bifurcation point ($\varepsilon_{HB2}$). Two points at two different $\varepsilon$ values are marked by solid circles ( red $\rightarrow$ $\varepsilon=0.5$ and green $\rightarrow$ $\varepsilon=1.0$). For these two marked points, the time series and phase space plots are shown in Fig.\ref{classical_tsps}.
%=====================bf
%=====================ef

\subsection{Scalar coupling: Classical AD and OD}
\label{scalaradod}
The amplitude equation corresponding to Eq.~\eqref{classical_eqn} under the scalar coupling is given by (by considering $\ve_1=\ve$ and $\ve_2=0$)
\begin{equation}
\label{mfd_amp}
\begin{split}
\dot{\alpha_j}&=-i\omega\alpha_j+(\frac{k_1}{2}-k_2|\alpha_j|^2)\alpha_j+\frac{\varepsilon}{2}\left(\frac{q}{2}\sum_{m=1}^{2}\alpha_m-\alpha_j\right)\\
&+\frac{\varepsilon}{2}\left(\frac{q}{2}\sum_{m=1}^{2}\alpha^*_m-\alpha^*_j\right).
\end{split}
\end{equation}
Eq.~\eqref{mfd_amp} has the following fixed points: the trivial fixed point $\mathcal{F}_{HSS} \equiv (0, 0, 0, 0)$, and additionally a coupling dependent nontrivial fixed point $\mathcal{F}_{IHSS} \equiv$ (${x_1}^\ast$, ${y_1}^\ast$, $-{x_1}^\ast$, $-{y_1}^\ast$) where ${x_1}^\ast = -\frac{\omega {y_1}^\ast}{{\omega}^2 + \varepsilon {{y_1}^\ast}^2}$ and ${y_1}^\ast = \sqrt {\frac{(\varepsilon - 2{\omega}^2) + \sqrt{{\varepsilon}^2 - 4{\omega}^2}}{2\varepsilon}}$. The system shows a transition from oscillatory state to amplitude death state through an inverse Hopf bifurcation at $\varepsilon_{\mbox{HB,s}}=\frac{k_1}{(1-q)}$ \cite{tanpre1,tanpre2}. An interesting transition from AD to OD state occurs through a {\it symmetry-breaking} pitchfork bifurcation at $\epsilon_{\mbox{PB}}=\frac{\omega^2}{k_1}$ \cite{tanpre1,tanpre2}. This transition from AD to OD is analogous to the Turing-type bifurcation in spatially extended system \cite{kosprl}. Figure~\ref{classical_bifur}(b) shows the corresponding bifurcation diagram for an exemplary parameter set ($\omega=2$, $k_1=1$, $k_2=0.2$, and $q=0.2$). Here we can see that unlike AD state, in the OD state, two branches of non zero steady IHSS emerges, which are placed symmetrically around the origin: $x_1=-x_2$ and $y_1=-y_2$. 
%It is noteworthy that in the OD state $y_{1,2}$ is greater than $x_{1,2}$ at a fixed $\varepsilon$. 

Our main aim in this work is to explore the quantum manifestation of the above mentioned classical results. In particular, we try to reveal the quantum mechanical analog of the symmetry-breaking OD state.

%{\cc \texttt{give bifurcation diagram of classical scalar (AD-OD) and vecor coupling (AD)}}

%==========================(VECTOR COUPLING)==============================

\section{Quantum vdP oscillators under nonscalar coupling: Quantum AD}\label{sec:qad} 
\subsection{Pure quantum oscillators}
The quantum master equation of two nonscalar mean-field diffusively coupled identical quantum van der Pol oscillators is given by,
%~~~~~~~~~~~~~~~~~~~~~~~~~~~~~~~~~~~~~~~~~~~~b_eqn
\begin{equation}
\label{mfd_master-vec}
\begin{split}
\dot{\rho}&=\sum_{j=1}^2\left(-i[H_j,\rho]+k_1\mathcal{D}[a_j^\dag](\rho)+k_2\mathcal{D}[{a_j}^2](\rho)\right)\\
&+q\varepsilon\mathcal{D}[(a_1+a_2)^\dag](\rho)+2\varepsilon\sum_{j=1}^2\mathcal{D}[a_j](\rho),
\end{split}
\end{equation}
%~~~~~~~~~~~~~~~~~~~~~~~~~~~~~~~~~~~~~~~~~~~~e_eqn
where, $H_j=\omega a_j^\dag a_j$ and $a_j$ ($a_j^\dag$) is the annihilation (creation) operator corresponding to the $j$-th oscillator.  
%%%%%%Correspondence between Master equation and Amplitude equation}
In the classical limit ($k_1>k_2$) the master equation \eqref{mfd_master-vec} is equivalent to the classical amplitude equation \eqref{amp_eqn-vec} through the following relation: $\dot{\braket{a}}=\mbox{Tr}(\dot \rho a)$.

\begin{figure}
%\vspace{0.4cm}
\includegraphics[width=.48\textwidth]{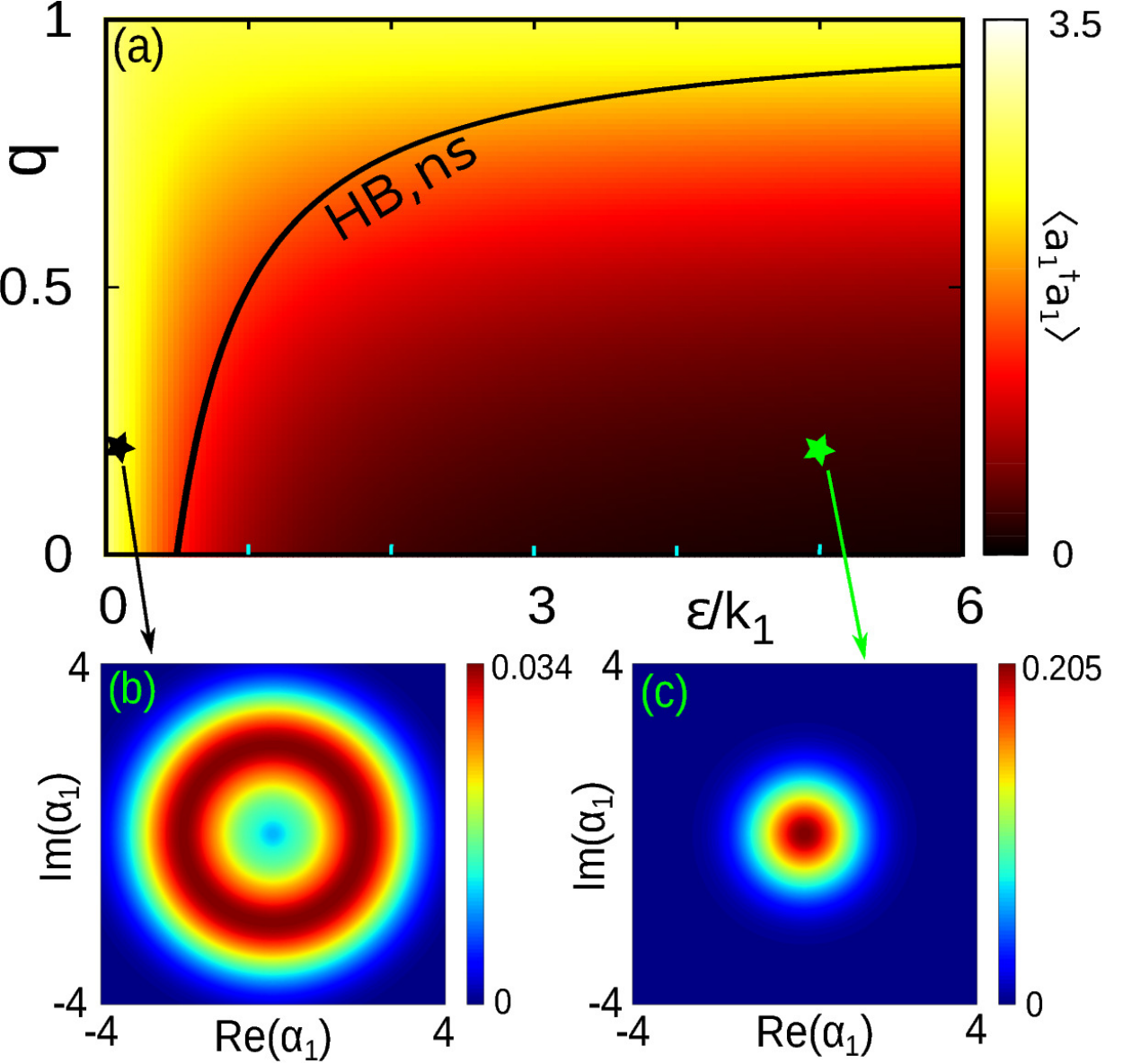}
\caption{(a) Two parameter diagram in the $\ve/k_1-q$ space showing the mean phonon number $\braket{a_1^\dag a_1}$ ($=\braket{a_2^\dag a_2}$) of both the oscillators. The solid line is the classical Hopf bifurcation curve (HB,ns). The steady state Wigner functions at (b) $\ve/k_1=0.1$ showing limit cycle oscillation and (c) $\ve/k_1=5$ for $q=0.2$ showing quantum AD. Other parameters are $k_1=1$, $k_2=0.2$ and $\omega=2$.}
\label{qm-2p}
\end{figure}

We numerically solve the master equation \eqref{mfd_master-vec} using \texttt{QuTiP} \cite{qutip}. To visualize and understand the system dynamics we employ the Wigner function representation in phase space since it provides a reliable representation of the quantum  dynamical states. Moreover, Wigner function is also an experimentally observable quasi-probability distribution function that makes it accessible \cite{wigner}. 
We computed the mean phonon number $\braket{a_{1}^\dag a_{1}}$ ($=\braket{a_{2}^\dag a_{2}}$) and plot them in the $\ve/k_1-q$ parameter space (since the oscillators are identical, both have the same mean phonon numbers). Figure \ref{qm-2p}(a) shows this in color map. The solid line indicates the Hopf bifurcation curve (HB,ns) obtained classically: below the Hopf curve the classical AD occurs. It is interesting to note that the mean phonon number also decreases appreciably under the Hopf bifurcation curve and due to the hindrance from quantum noise it does not reach zero but shows a moderate collapse in the oscillation. However, this moderate collapse is stronger than the noisy classical oscillators (discussed in the next subsection). The corresponding steady state Wigner function of two representative points are shown in Figs.~\ref{qm-2p}(b c): Fig.~\ref{qm-2p}(b) demonstrates the oscillatory behavior for $\ve/k_1=0.1, q=0.2$ and Fig.~\ref{qm-2p}(c) shows the quantum AD for $\ve/k_1=5, q=0.2$. 

%We also plot the Mandel $Q_M$ number defined as:$\braket{(n_j-\braket{n_j})^2}/\braket{n_j}-1$. This will give an estimate of the relation between the mean phonon number and the phonon statistics. From Fig.~\ref{qm-2p}(d) it is found that $Q_M$ is always positive, which is consistent with the fact that the Wigner function is always positive. 

\begin{figure}
\vspace{0.4cm}
\includegraphics[width=.46\textwidth]{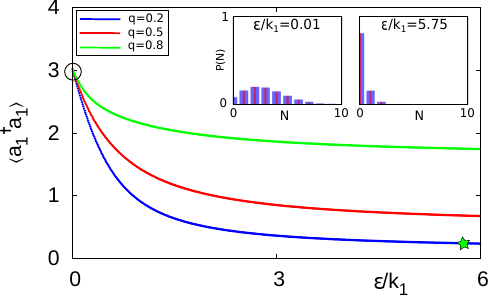}
\caption{(a) The steady state mean phonon number $\braket{a_{1}^\dag a_{1}}$ ($=\braket{a_{2}^\dag a_{2}}$) for three different values of mean-field density ($q=0.2, 0.5, 0.8$). Inset shows the occupation of Fock levels for oscillatory state at $\varepsilon/k_1=0.01$ (left panel) and quantum AD at $\varepsilon/k_1=5.75$ (right panel) with $q=0.2$. Other parameters are $k_1=1$, $k_2=0.2$ and $\omega=2$.} 
\label{quantum}
\end{figure}

The variation of the mean phonon number ($\braket{a_{1}^\dag a_{1}}$) with three different values of $q$ is shown in Fig.~\ref{quantum}. It is found that quantum AD occurs more effectively in the lower $q$ values. This is due to the fact that a lower (higher) $q$ imposes stronger (weaker) dissipation in the coupled system. 
%=====================bf
%=====================ef
%=====================bf
%=====================ef
In Fig.~\ref{quantum}, the probability of occupation of the Fock levels are shown in the insets for two coupling strengths at $q=0.2$: for $\varepsilon/k_1=0.01$ (left inset) the system shows oscillation and for $\varepsilon/k_1=5.75$ (right inset) quantum AD appears. It is clear that in the quantum AD state the occupation near the quantum ground state is much more prominent.

We also explored the deep quantum region where strong nonlinear damping rate dominates the linear pumping rate ($k_2\gg k_1$) and got qualitatively similar results (not shown here). In the deep quantum regime only a few states are populated, which are near the quantum ground state: note that in the limit $k_2\rightarrow\infty$ the steady state density matrix is given by \cite{lee_prl} $\rho_{ss}=\frac{2}{3}|0\rangle\langle 0|+\frac{1}{3}|1\rangle\langle 1|$, i.e., the system oscillates between its two lowest lying energy levels. Therefore, the notion of quantum AD is not obvious in the deep quantum regime as the mean phonon number always remains very low irrespective of the coupling conditions.

%****************************************************************
%***********************(section_start)**************************
%****************************************************************

\subsection{Noisy classical model}
\label{semi-nonscalar}
For the proper understanding of quantum AD it is instructive to compare the results of the quantum system with the corresponding noisy classical model (or semiclassical model) \cite{qad1}. In the noisy classical model the classical dynamics is considered in the presence of a finite amount of noise whose intensity is equal to that of the quantum noise. To evaluate the amount of quantum noise intensity, a stochastic differential equation is derived from the quantum master equation following \cite{qad1}. 
For this, the quantum master equation \eqref{mfd_master-vec} is represented in phase space using partial differential equation of Wigner distribution function ($W(\bm{\alpha})$) \cite{carmichael}. 
%~~~~~~~~~~~~~~~~~~~~~~~~~~~~~~~~~~~~~~~~~~~~b_eqn
\begin{equation}
\label{diff_eqn_w}
\begin{split}
\partial_t{W(\bm{\alpha})}&=\sum_{j=1}^2\left[-\left(\frac{\partial}{\partial \alpha _j}\mu _{\alpha _j}+c.c.\right) \right. \\
&+ \left. \frac{1}{2}\left(\frac{\partial ^2}{\partial \alpha _j \partial {\alpha_j}^*}D_{\alpha _j {\alpha_j}^*}+\frac{\partial ^2}{\partial \alpha _j \partial {\alpha_{j'}}^*}D_{\alpha _j {\alpha_{j'}}^*} \right) \right. \\
&+ \left. \frac{k_2}{4}\left(\frac{\partial ^3}{\partial {\alpha_j}^* \partial {\alpha_j}^2}\alpha _j+c.c\right) \right]W(\bm{\alpha}),
\end{split}
\end{equation}
where the elements of the drift vector ($\bm{\mu}$) are: $$\mu _{\alpha _j}=\left[-i\omega+\frac{k_1}{2}-k_2(|\alpha _j|^2-1)-\left(\varepsilon-\frac{\varepsilon q}{2}\right)\right]\alpha _j+\frac{\varepsilon q}{2}\alpha _{j'},$$
and the elements of the diffusion matrix $\bm{D}$ are:
$$D_{\alpha _j {\alpha_j}^*}=k_1+2k_2(2|\alpha _j|^2-1)+\varepsilon q+2\varepsilon, D_{\alpha _j {\alpha_{j'}}^*}=\varepsilon q.$$
with $j=1,2$, $j'=1,2$ and $j\neq j'$. In weak nonlinear regime ($k_2\ll k_1$), Eq.\ref{diff_eqn_w} reduces to the Fokker-Planck equation, which is given by 
%~~~~~~~~~~~~~~~~~~~~~~~~~~~~~~~~~~~~~~~~~~~~b_eqn
\begin{equation}
\label{fp}
\begin{split}
\partial_t{W}(\textbf{X})&=\sum_{j=1}^2\left[-\left(\frac{\partial}{\partial x_j}\mu _{x_j}+\frac{\partial}{\partial y_j}\mu _{y_j}\right) \right) \\
&+ \left. \frac{1}{2}\left(\frac{\partial ^2}{\partial x_j \partial x_j}D_{x_j x_j}+\frac{\partial ^2}{\partial y_j \partial y_j}D_{y_j y_j} \right. \right. \\
&+ \left. \left. \frac{\partial ^2}{\partial x_j \partial x_{j'}}D_{x_j x_{j'}}+\frac{\partial ^2}{\partial y_j \partial y_{j'}}D_{y_j y_{j'}}  \right) \right]W(\textbf{X}),
\end{split}
\end{equation}
%~~~~~~~~~~~~~~~~~~~~~~~~~~~~~~~~~~~~~~~~~~~~e_eqn
where $\textbf{X}=(x_1, y_1, x_2, y_2)$. The elements of drift vector are,
%~~~~~~~~~~~~~~~~~~~~~~~~~~~~~~~~~~~~~~~~~~~~b_eqn
\begin{subequations}
\label{drift}
\begin{align}
\begin{split}
\mu _{x_j}&=\omega y_j + \left[\frac{k_1}{2}-k_2({x_j}^2+{y_j}^2-1) \right. \\
&- \left. \left(\varepsilon-\frac{\varepsilon q}{2}\right)\right]x_j+\frac{\varepsilon q}{2}x_{j'},
\end{split} \\
\begin{split}
\mu _{y_j}&=-\omega x_j + \left[\frac{k_1}{2}-k_2({x_j}^2+{y_j}^2-1) \right. \\
&- \left. \left(\varepsilon-\frac{\varepsilon q}{2}\right)\right]y_j+\frac{\varepsilon q}{2}y_{j'}.
\end{split}
\end{align}
\end{subequations}
%~~~~~~~~~~~~~~~~~~~~~~~~~~~~~~~~~~~~~~~~~~~~e_eqn
The diffusion matrix has the following form,
%~~~~~~~~~~~~~~~~~~~~~~~~~~~~~~~~~~~~~~~~~~~~b_eqn
\begin{equation}\label{diffusion_mat} 
{\bm{D}}=\frac{1}{2}\left(\begin{array}{cccc} \nu _1 & 0 & \frac{\varepsilon q}{2} & 0 \\
0 & \nu _1 & 0 & \frac{\varepsilon q}{2}\\
\frac{\varepsilon q}{2} & 0 & \nu _2 & 0 \\
 0 & \frac{\varepsilon q}{2} & 0 & \nu _2  \end{array}\right).
\end{equation}
%~~~~~~~~~~~~~~~~~~~~~~~~~~~~~~~~~~~~~~~~~~~~e_eqn
where $\nu _j=\frac{k_1}{2}+k_2[2({x_j}^2+{y_j}^2)-1]+\frac{\varepsilon q}{2}+\varepsilon$.
From Eq.\eqref{fp}, the following stochastic differential equation can be derived,
%~~~~~~~~~~~~~~~~~~~~~~~~~~~~~~~~~~~~~~~~~~~~b_eqn
%\begin{subequations}
\begin{equation}\label{sde}
d\textbf{X}=\bm{\mu}dt+\bm{\sigma} d\textbf{W}_t,
\end{equation}
%\end{subequations}
%~~~~~~~~~~~~~~~~~~~~~~~~~~~~~~~~~~~~~~~~~~~~e_eqn
where $\bm{\sigma}$ is the noise strength and $d\textbf{W}_t$ is the Wiener increment. As the diffusion matrix $\bm{D}$ (given in Eq.\eqref{diffusion_mat}) is symmetric, we can analytically derive $\bm{\sigma}$ from it. The diagonal form of $\bm{D}$ is given by: $\bm{D}_{diag}=\textbf{U}^{-1}\bm{D}\textbf{U}=diag(\lambda_- \lambda_- \lambda_+ \lambda_+)$. Here $\lambda_{\pm}=\frac{1}{4}\left[\nu _1+\nu _2\pm\sqrt{(\nu _1-\nu _2)^2+(\varepsilon q)^2}\right]$ and $\textbf{U}$ has the following form:
%~~~~~~~~~~~~~~~~~~~~~~~~~~~~~~~~~~~~~~~~~~~~b_eqn
\begin{align}\label{u_mat}
{\mbox {\bf U}}&=\left(\begin{array}{cccc} 0 & u_- & 0 & u_+ \\
u_- & 0 & u_+ & 0\\
0 & 1 & 0 & 1 \\
1 & 0 & 1 & 0  \end{array}\right),
\end{align}
%~~~~~~~~~~~~~~~~~~~~~~~~~~~~~~~~~~~~~~~~~~~~e_eqn
where $u_{\pm}=\frac{1}{\varepsilon q}\left[\nu _1-\nu _2\pm\sqrt{(\nu _1-\nu _2)^2+(\varepsilon q)^2}\right]$. Now, $\bm{\sigma}$ matrix can be evaluated from the equation $\bm{\sigma}=\textbf{U}\sqrt{\bm{D}_{diag}}\textbf{U}^{-1}$ and it has the following form.
%~~~~~~~~~~~~~~~~~~~~~~~~~~~~~~~~~~~~~~~~~~~~b_eqn
\begin{align}\label{sigma_mat}
{{\bm{\sigma}}}&=\left(\begin{array}{cccc} \sigma_1 & 0 & \sigma_3 & 0 \\
0 & \sigma_1 & 0 & \sigma_3\\
\sigma_3 & 0 & \sigma_2 & 0 \\
0 & \sigma_3 & 0 & \sigma_2  \end{array}\right),
\end{align}
%~~~~~~~~~~~~~~~~~~~~~~~~~~~~~~~~~~~~~~~~~~~~e_eqn
where $\sigma_1=\frac{u_+\sqrt{\lambda_+}-u_-\sqrt{\lambda_-}}{u_+-u_-}$, $\sigma_2=\frac{u_+\sqrt{\lambda_-}-u_-\sqrt{\lambda_+}}{u_+-u_-}$ and $\sigma_3=\frac{\sqrt{\lambda_+}-\sqrt{\lambda_-}}{u_+-u_-}$.

By solving the stochastic differential equation (Eq.~\eqref{sde}) (using JiTCSDE module in Python \cite{jitcode}), we compute the ensemble average of the squared steady-state amplitude of the first oscillator ($\overline{{|\alpha_1|_{nc}}^2}$), averaged over 1000 realizations, starting from random initial conditions. 
%=====================bf
\begin{figure}
%\vspace{0.4cm}
\includegraphics[width=.48\textwidth]{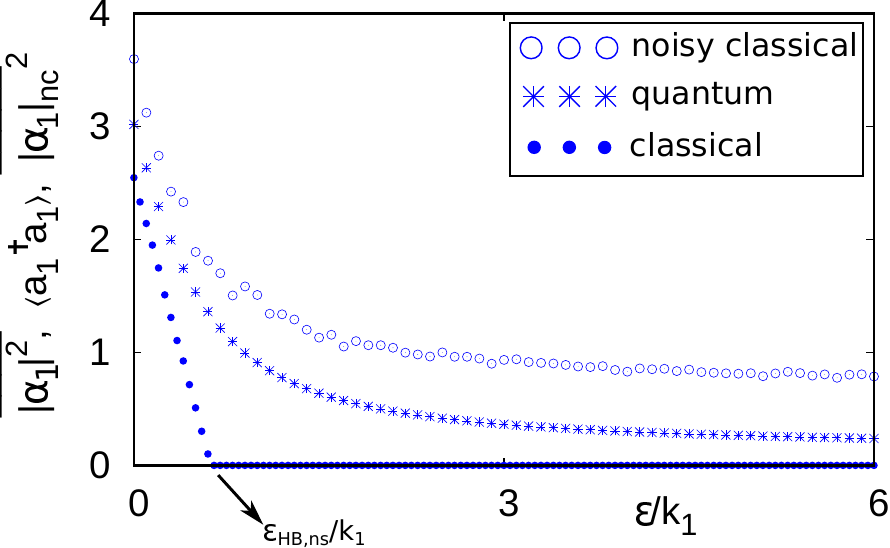}
\caption{{\bf Nonscalar coupling:} Comparison of the classical, quantum, and semiclassical results. At $q=0.2$, the average amplitude from the classical model ($\overline{{|\alpha _1|}^2}$), mean phonon number from the quantum model ($\braket{a_1^\dag a_1}$) and the averaged amplitude from the noisy classical model ($\overline{{|\alpha _1|_{nc}}^2}$) of the first oscillator plotted together with coupling strength. Other parameters are $k_1=1$, $k_2=0.2$, and $\omega=2$.}
\label{all_models}
\end{figure}
%=====================ef
To compare the scenarios of oscillation collapse for each model, in Fig.~\ref{all_models} we plot the averaged amplitude of classical model  ($\overline{{|\alpha _1|}^2}$), that of the noisy classical model ($\overline{{|\alpha _1|_{nc}}^2}$) and the mean phonon number of the quantum model ($\braket{a_1^\dag a_1}$) of the first oscillator with the coupling parameter. 
%In the case of classical and noisy classical models, the amplitude is averaged over 1000 real time of evolution. 
The averaged classical amplitude shows an abrupt jump from oscillatory state to death state at $\varepsilon _{HB,ns}$. Whereas, the mean phonon number and the averaged amplitude of noisy classical model do not show a zero-amplitude death state, rather they show a significant decrement in amplitude. It can be seen that the mean phonon number is always lesser than the average amplitude of the noisy classical model. Therefor the quantum AD lies in between the classical AD and the AD in the noisy classical model.
%decrement in the mean phonon number is the leads to the oscillation collapse in coupled identical quantum van der Pol oscillators.

\section{Quantum vdP oscillators under scalar coupling: Quantum OD state} \label{sec:qod}
The quantum master equation of two coupled identical quantum van der Pol oscillators under scalar coupling is given by
\begin{equation}
\label{mfd_master}
\begin{split}
\dot{\rho}&=-i\bigg[\omega (a_1^\dag a_1+a_2^\dag a_2)+\frac{i\epsilon}{4}\big(q(a_1^\dag a_2^\dag-a_{1}a_{2})\\
&+(\frac{q}{2}-1)({a_1^\dag}^2+{a_2^\dag}^2-{a_1}^2-{a_2}^2)\big),\rho\bigg]\\
&+k_1\sum_{j=1}^2\mathcal{D}[a_j^\dag](\rho)+k_2\sum_{j=1}^2\mathcal{D}[{a_j}^2](\rho)\\
&+\frac{q\epsilon}{2}\mathcal{D}[(a_1+a_2)^\dag](\rho)+\epsilon\sum_{j=1}^2\mathcal{D}[a_j](\rho).
\end{split}
\end{equation}
In the classical limit, \eqref{mfd_master} gives the classical amplitude equation \eqref{mfd_amp} using $\dot{\braket{a}}=\mbox{Tr}(\dot\rho a)$. 

\begin{figure}
\includegraphics[width=.48\textwidth]{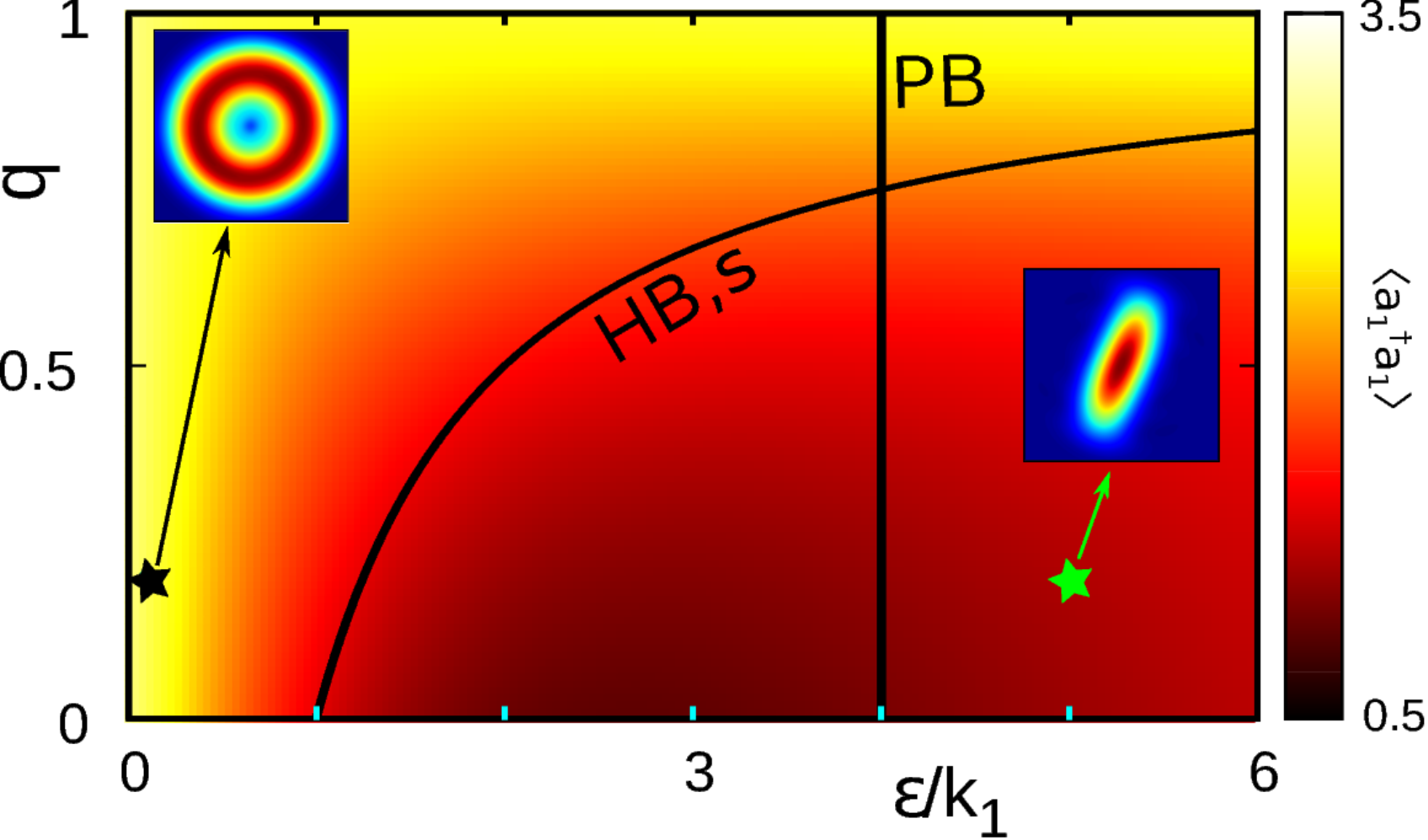}
\caption{{\bf Scalar coupling:} Two parameter phase diagram of the mean phonon number $\braket{a_1^\dag a_1}$ for $(k_1, k_2)=(1,0.2)$. Insets show Wigner function in the oscillatory state for $\ve/k_1=0.1$ and $q=0.2$  and {\it squeezed} quantum AD for $\ve/k_1=5$ and $q=0.2$. In the quantum AD state note the presence of squeezing in the quadrature space. In the Wigner plots axes and scales are identical to Fig.~\ref{qm-2p}. Other parameter: $\omega=2$.}
\label{2pscalar}
\end{figure}
%=====================ef
\begin{figure}
\includegraphics[width=.45\textwidth]{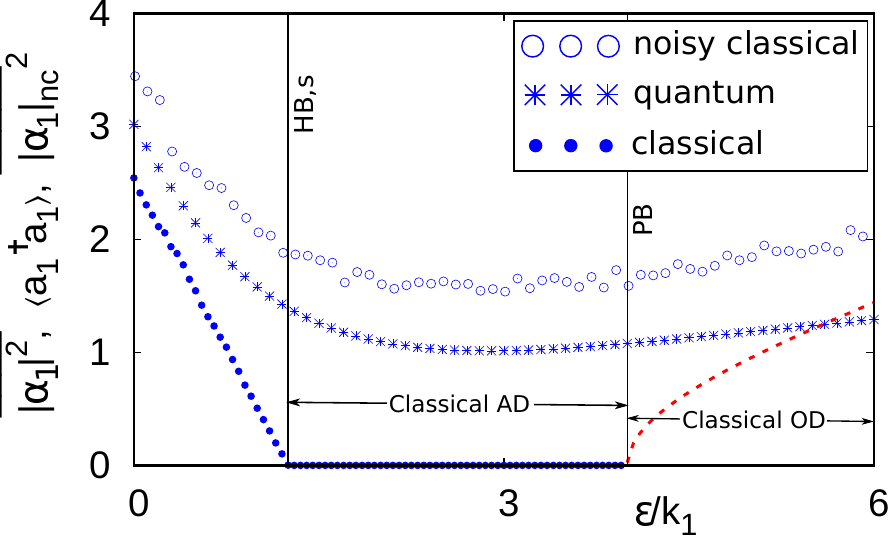}
\caption{{\bf Scalar coupling:} Comparison of the classical, quantum, and semiclassical results. Plots of the average amplitude from the classical model ($\overline{{|\alpha _1|}^2}$), mean phonon number from the quantum model ($\braket{a_1^\dag a_1}$) and the averaged amplitude from the noisy classical model ($\overline{{|\alpha _1|_{nc}}^2}$) of the first oscillator at $q=0.2$. Red dashed line represents the shift of the stable inhomogeneous fixed points from the origin in the OD state. Other parameters are $k_1=1$, $k_2=0.2$ and $\omega=2$.}
\label{semi-scalar}
\end{figure}

We solve the master equation \eqref{mfd_master} numerically using \texttt{QuTiP} \cite{qutip}. 
%Since we are interested to explore the quenching dynamics in the deep quantum region, therefore, we choose $\kappa_1=1$ and $\kappa_2=10$. 
At first, similar to the nonscalar coupling case of the the previous section, we consider $k_1=1$ and $k_2=0.2$. The results are summarized in Fig.~\ref{2pscalar}: it shows the mean phonon number $\braket{a_1^\dag a_1}$ ($=\braket{a_2^\dag a_2}$) along with the classical Hopf and pitchfork bifurcation curves in the $\ve/k_1-q$ parameter space (classical bifurcation curves are drawn using the expressions derived in Sec.~\ref{scalaradod}). The insets in Fig.~\ref{2pscalar} show the Wigner function in phase space of oscillatory and quantum AD state for $\ve/k_1=0.1$ and $\ve/k_1=5$, respectively at fixed $q=0.2$. 
%It is evident from the that in the quantum amplitude death state intrinsic quantum noise forbids the complete quenching of the dynamics. 
An interesting observation from the Wigner function plot of the quantum AD is the presence of squeezing in the quadrature space. The squeezing gets stronger with increasing $\ve$. This may be the direct reflection of the classical case, where under {\it scalar coupling} $x<y$ (for $x,y\ne 0$) for a nonzero coupling strength. This type of squeezing is not present in the case of nonscalar coupling or in diffusive coupling \cite{qad1,qad2} as coupling is symmetric there with respect to all the variables.

%=====================bf
\begin{figure*}
\includegraphics[width=.95\textwidth]{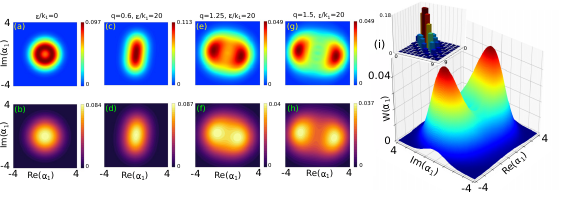}
\caption{{\bf Quantum manifestation of oscillation death (OD):}  Deep quantum region ($k_1=1$ and $k_2=3$).  (a, c, e, g) Wigner function (b, d, f, h) Husimi function. (a, b) Limit cycle oscillation in uncoupled oscillators ($\ve/k_1=0$) (c, d) Squeezed qantum AD for $\ve/k_1=20$, $q=0.6$. Quantum OD state for (e, f) $q=1.25$ and (g, h) $q=1.5$: note the emergence of inhomogeneous steady states. (i) Three dimensional representation of (g). It clearly shows two separated lobes in the phase space. Inset shows the histogram of the real part  of elements of the density matrix: note the presence of coherence terms. Other parameter: $\omega=2$.}
\label{odcat}
\end{figure*}
%=====================ef

In Fig.~\ref{2pscalar} classical AD occurs below the Hopf bifurcation curve (HB,s) and left to the PB line and classical OD occurs below the HB,s curve right to the PB line. Although, the occurrence of classical and quantum AD agrees with each other in the parameter space, surprisingly, for no value of $\varepsilon$ and $q$, inhomogeneous steady states (OD) are observed; squeezed quantum AD appears even beyond the PB line (below the HB,s curve). However, a slight increase in mean phonon number $\braket{a_1^\dag a_1}$ is observed beyond the classical PB line. For better understanding of the fact we study the noisy classical model using the formalism equivalent to Sec.~\ref{semi-nonscalar}. Figure~\ref{semi-scalar} shows the plots of average amplitude and mean phonon number in classical, semiclassical and quantum oscillators at a fixed $q=0.2$. Upto the PB line, the quantum AD scenario of Fig.~\ref{semi-scalar} qualitatively matches with Fig.~\ref{all_models} of the nonscalar case. Beyond the PB line, in the classical case, oscillation ceases, and nonzero inhomogeneous fixed points are created that are shifted from the origin: the (red) dashed line shows the amount of shift of the fixed points in the classical OD state. However, in the quantum as well as noisy classical cases, noise tends to homogenize the steady states around the origin. As a result no quantum OD is observed here, rather, it results in a slight increase in the mean phonon number (in quantum case) or average amplitude (noisy classical case).

%leads to increased squeezing and an increase in $q$ leads to larger mean phonon number. {\cc results of semiclassical treatment?}
%=====================bf

Next, we search for any possible symmetry-breaking dynamics in the deep quantum regime ($k_2>k_1$). Following \cite{squeezing} we choose $k_1=1$ and $k_2=3$. In this regime only a few Fock states are populated (near the quantum ground state) and quantum noise becomes much more prominent. Therefore, it is inconclusive to distinguish between the oscillatory state and the quantum AD state based on the mean phonon number. However, qualitative changes in the Wigner function provides distinction between them. Figure~\ref{odcat}(a) shows the Wigner function representation of oscillation in the uncoupled case ($\ve=0$) and Fig.~\ref{odcat}(c) shows that for quantum AD at $\ve/k_1=20$ and $q=0.6$ [Fig.~\ref{odcat}(b) and (d) show the respective Husimi function \cite{husimi}]. While in the oscillatory case the origin shows a dip in the value of Wigner function, in the quantum AD state it shows a peak. 
%Here also, $0\le q\le 1$ we do not observe any inhomogeneous steady state.  

%{\cc is it possible to give a 2 parameter plot showing OD?}
With further increase in $q$, at a moderate $\ve$ we observe an interesting symmetry-breaking bifurcation that governs the creation of inhomogeneous steady states, i.e., the quantum oscillation death state.  
The quantum OD state emerges as the Wigner function (and therefore, the Husimi function) bifurcates into two separated lobes in the phase space. 
Figures~\ref{odcat}(e--h) demonstrate the quantum OD state in the phase space using Wigner function (upper row) and Husimi function (bottom row) for two representative values of $q$: Figs.~\ref{odcat}(e, f) are for $q=1.25$ and Figs.~\ref{odcat} (g, h) are for $q=1.5$. One can observe that the probability density is concentrated in the two lobes. The separation between the two lobes increases with increasing $q$. The three dimensional plot of Fig.~\ref{odcat}(g) as shown in Fig.~\ref{odcat}(i) adds more clarity to the occurrence of symmetry-breaking bifurcation and creation of the quantum OD state. 
%We find that it is the quantum manifestation of the oscillation death state. 
We observed that the Wigner function in the OD state is not just two lobes separated in the phase space, which is nothing but classical representation of probability of two possible outcomes \cite{pracat}, however, in this case quantum interference terms appear in the middle and exhibits symmetry-breaking inhomogeneous steady states. This fact is also verified by the non zero coherence terms in the density matrix of this state: the inset of Fig.~\ref{odcat}(i) shows the histogram of the real part of the elements of the density matrix that exhibits the presence of off-diagonal terms in the density matrix. %Therefore, the ``quantum OD" state is indeed a quantum manifestation of classical OD state.
Since the quantum OD state occurs in the deep quantum regime, we can not draw a one to one correspondence with the classical OD state as now the classical amplitude equation \eqref{mfd_amp} and the quantum master equation \eqref{mfd_master} are no longer exactly equivalent.
%To support our result we use the measure of mixedness. It is shown that coherence of the steady state density matrix in the OD state are non zero. A (classical) statistical mixture would result in zero off-diagonal terms. 
It is noteworthy that in Ref.~\cite{squeezing} a symmetry-breaking bifurcation in the Wigner distribution function occurs in a squeezing driven van der Pol oscillator. However, that does not resemble an OD state as it occurs in a single driven oscillator. In our case the symmetry-breaking bifurcation occurs due to the coupled interaction of two oscillators, therefore, the notion of emergent dynamics is applicable here. 

\begin{figure}
\includegraphics[width=.45\textwidth]{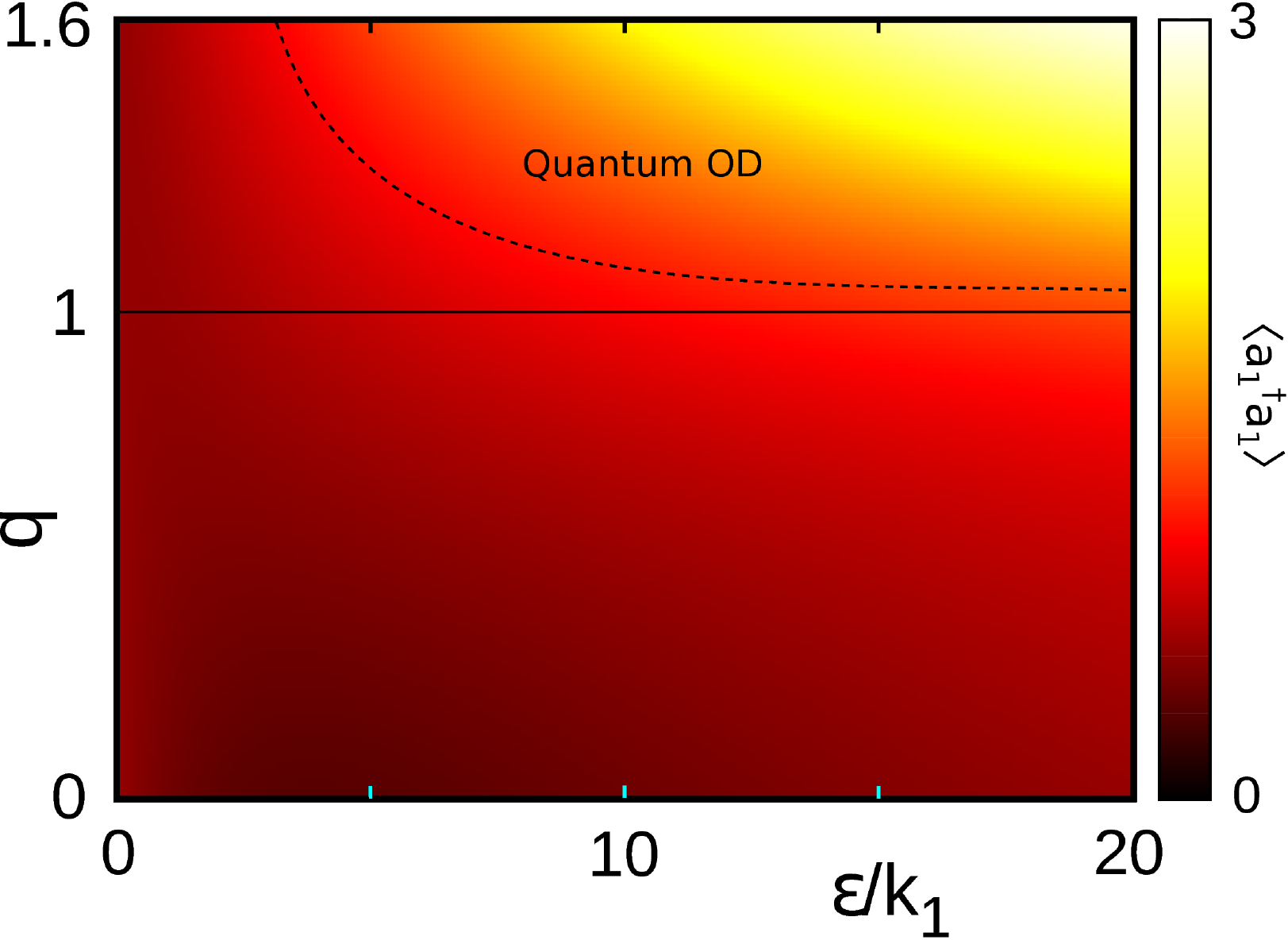}
\caption{{\bf Deep quantum region $k_1=1$ and $k_2=3$:} Mean phonon number $\braket{a_1^\dag a_1}$ in the two parameter space. Quantum OD occurs above the dashed line. Solid horizontal line indicates $q=1$. $\omega=2$.}
\label{2pod}
\end{figure}

Finally, in the deep quantum regime we tried to map the zone of occurrence of the quantum OD state in the $\ve/k_1-q$ space. Figure~\ref{2pod} shows the mean phonon number in the $\ve/k_1-q$ space ($k_1=1, k_2=3$): quantum OD emerges above the dashed line which is plotted by visual inspection of the bifurcation of Wigner function. In the quantum OD state one observes a drastic increase in the mean phonon number, which resembles the fact that in classical OD, the inhomogeneous fixed points have non zero values. However, an exact demarcation of the quantum OD in the parameter space is difficult in the absence of any quantitative measure of this state.  

At this point it is important to raise the issue of quantum mechanical analog of classical Turing-type bifurcation. As discussed earlier, in Ref.~\cite{kosprl} the transition from AD to OD through a symmetry-breaking bifurcation was established as equivalent to the Turing-type bifurcation of spatially extended system. In the present study also, in the deep quantum regime we get quantum AD and OD in the same system. Moreover, we notice a symmetry-breaking transition from the squeezed AD state to the quantum OD state with increasing $q$ (cf. Fig.~\ref{odcat}): this may be thought of as the ``quantum" analog of the Turing-type bifurcation. However, since quantum OD occurs in the deep quantum regime only, therefore, the presence of strong quantum noise makes it  difficult to distinguish quantum AD from oscillations and to identify the exact route of transition from quantum AD to quantum OD. Therefore, more quantitative measures are required to draw any strong conclusions regarding this.

\section{Conclusions}
\label{sec:con}
In this paper we have studied the quantum mechanical manifestations of oscillation suppression states, namely the amplitude death and the oscillation death states in two mean-field diffusively coupled identical quantum van der Pol oscillators. Our study has unraveled two questions that we asked in the beginning of this paper. First, {\it identical} quantum oscillators can exhibit two types of quantum amplitude death states, namely squeezed and non squeezed quantum AD. Second, oscillation death state indeed appears in coupled quantum oscillators; it is manifested in the deep quantum regime as the creation of inhomogeneous steady states due to symmetry-breaking bifurcation in Wigner function and Husimi function. Moreover, our results hint at the occurrence of the quantum analog of the Turing-type bifurcation.  

First we have shown that under nonscalar coupling quantum AD state appears in its  non squeezed form.
With the scalar coupling we observed a {\it squeezed} quantum AD state, which is unlike nonscalar or diffusive coupling induced quantum AD state \cite{qad1,qad2}.
In the higher excitation regime (i.e., outside the deep quantum zone) the quantum AD has a one to one correspondence with the classical and semiclassical results. However, in the deep quantum regime the notion of quantum AD is not obvious because in this regime only a few Fock levels are populated around the quantum mechanical ground state.  
%It has been noticed that the oscillation collapse is more intense in quantum model than noisy classical model. 
%For the first time we report the quantum equivalent of the oscillation death (OD) state. 
%We have shown that the quantum OD state does not appear outside the deep quantum regime. 

In the deep quantum regime with high mean-field density we have discovered a quantum OD state that emerges as the consequence of symmetry-breaking bifurcation in the Wigner distribution function. To the best of our knowledge this is the first instance where the quantum equivalence of the OD state has been observed. Since this state is exhibited in the deep quantum regime, therefore, a one to one correspondence with the classical OD state is not possible. However, both quantum and classical OD share two common features. First, they appear under the scalar coupling, and second, their manifestation in the phase space is equivalent, viz., the appearance of inhomogeneous steady states in the phase space. 
%Classical OD and quantum OD differs from their dependence on the coupling parameters: while increasing coupling strength (for a fixed mean-field density) gives rise to classical OD, quantum OD occurs for increasing  mean-field density (for a fixed coupling strength {\cc QOD independent of $\ve$?}).  
Since one of our main goals in this paper is the observation of OD in quantum regime, therefore, we restrict our study to two coupled oscillators. In the case of more than two oscillators, multicluster OD may appear \cite{kosprep} and the identification of the same in the quantum regime may become illusive.

With the advancement of experimental techniques we believe that the present coupling schemes can be realized experimentally, e.g., using the  ion trap \cite{lee_prl,expt-ion}  and ``membrane-in-the-middle" experimental set up \cite{expt-mem}. Quantum amplitude death is thought to be an efficient mean of cavity cooling \cite{qad1}; since in the present coupling scheme no parameter mismatch is required and one has two control parameters --- coupling strength and density of mean-field, therefore, we believe that the present scheme offers a more flexible option for cooling. Further, since a strong squeezing appears in the quantum AD state under the scalar coupling, therefore, generation of the squeezed state in coupled oscillators and its possible real life applications can be explored further \cite{sq-app2,sq-app3}.  On the other hand, OD is generally thought of as the underlying mechanism of cellular differentiation and other symmetry breaking phenomena in biological systems \cite{kosprep}, however,  we have to figure out the exact implication of the quantum OD state in real quantum systems. Only then we will be able to identify the application potentiality of quantum OD in quantum technology.
% of OD in quantum regime.   {\cc harnessing application in OD: Herald the engineering quantum systems}

The observation of quantum OD state opens up a myriads of scopes in the study of symmetry-breaking dynamics in the quantum regime. The shape of the Wigner function in the quantum OD state has a striking resemblance with that of the single-photon-subtracted two-mode states with vortex structure in quadrature space \cite{vortex} (see Chapter 4 of \cite{gsbook}); also, it shares some of the visual features of the squeezed Schr\"odinger cat (like) state \cite{knightbook,expt-cat}. However, unlike the vortex state and Schr\"odinger cat state in our system the Wigner function is always positive. Nevertheless, this visual resemblance calls for the further investigation.
%Therefore, the natural question arises about the connection between them. The exact characterization of the Schr\"odinger cat-like state will be an independent problem to study. The recent experimental observation of Schr\"odinger cat state in optical domain [cites] hints the possibility of observing the quantum OD in a real experimental set up.  
Our observation of quantum Turing-type bifurcation hints at the possibility of Turing pattern in quantum domain. However, a deep understanding of this scenario demands much more in-depth investigations.    
Recently, the symmetry-breaking partially synchronized states, namely the chimeras have been reported in quantum regime by \citet{schoell_qm}. The connection between the ``quantum" chimera states and the quantum OD state will be an interesting problem to study in a network of coupled quantum oscillators \cite{scholl_CD,tanCD}. 
\appendix
\section{Derivation of amplitude equation (Eq.~\ref{single_amp})}\label{app:A}
We consider the complex amplitude of the oscillator \eqref{vdp} as $\alpha=x+iy$. Therefore,
\begin{equation}
\begin{split}
\dot{\alpha}&=\dot{x}+i\dot{y},\\ 
&=\omega y-i\omega x+ik_1 y-8ik_2x^2y,\\
%&=-i\omega(x+iy)+ik_1\left(\frac{\alpha-\alpha^*}{2i}\right)\\
%&-8ik_2\left(\frac{\alpha+\alpha^*}{2}\right)^2\left(\frac{\alpha-\alpha^*}{2i}\right)\\
&=-i\omega \alpha + \frac{k_1}{2}(\alpha-\alpha^*)-k_2(\alpha+\alpha^*)^2(\alpha-\alpha^*).
\end{split}
\end{equation}
Using polar coordinate $\alpha=\eta e^{-i\phi}$ we get,
\begin{equation}
\label{etaphi}
\begin{split}
\dot{\eta}-i\eta\dot{\phi} &= -i\omega \eta + \frac{k_1}{2}\eta(1-e^{2i\phi})\\
&-k_2\eta^3 (e^{-i\phi}+e^{i\phi})^2(1-e^{2i\phi}).
%&= -i\omega \eta + \frac{k_1}{2}\eta(1-\cos{2\phi}-i\sin{2\phi})\\
%&-4k_2\eta^3\cos^2{\phi} (1-\cos{2\phi}-i\sin{2\phi})\\
\end{split}
\end{equation}
From Eq.\eqref{etaphi} we can extract equations for $\dot{\eta}$ and $\dot{\phi}$,
\begin{equation}
\begin{split}
\dot{\eta} &= \frac{k_1}{2}\eta(1-\cos{2\phi})-4k_2\eta^3\cos^2{\phi} (1-\cos{2\phi}),\\
\dot{\phi} &= \omega + \frac{k_1}{2}\sin{2\phi}-4k_2\eta^2\cos^2{\phi}\sin{2\phi}.\\
\end{split}
\end{equation}
Now at this point we apply the method of averaging. It can be done by directly averaging the equations of $\dot{\eta}$ and $\dot{\phi}$ over one time period $T=\frac{2\pi}{\omega}$ (for details see \cite{piko}, Chapter 7 and references therein). We get,
\begin{equation}
\begin{split}
\dot{\eta} &= (\frac{k_1}{2}-k_2\eta^2)\eta,\\
\dot{\phi} &= \omega.
\end{split}
\end{equation}
Putting these averaged values of $\dot{\eta}$ and $\dot{\phi}$ in the equation $\dot{\alpha}=e^{-i\phi}(\dot{\eta}-i\eta\dot{\phi})$ we get the following equation,
\begin{equation}
%\begin{split}
\dot{\alpha} =-i\omega\alpha + (\frac{k_1}{2} - k_2|\alpha|^2)\alpha.
%\end{split}
\end{equation}
This is the amplitude equation as given in Eq.\eqref{single_amp}.

%######################################## (Section) ###############################################
\section{Correspondence between master equation (Eq.~\ref{single_master}) and amplitude equation (Eq.~\ref{single_amp})}\label{app:B}
In quantum optics the average annihilation operator ($\braket{a}$) and the complex amplitude ($\alpha$) are equivalent \cite{knightbook}, i.e., $\braket{a}\equiv\alpha$. This property bridges the master equation and amplitude equation. Let us consider the master equation in the Lindblad form as $\dot{\rho}=-i[H,\rho]+\mathcal{D}[L](\rho)$. Now the average of any operator $\hat{O}$ is given by $\braket{\hat{O}}=\mbox{Tr}(\rho\hat{O})$. So the dynamical equation of $\braket{\hat{O}}$ is given by,
\begin{equation}
\label{odot}
\begin{split}
\frac{d\braket{\hat{O}}}{dt}&=\frac{d}{dt}\mbox{Tr}(\rho \hat{O}),\\
%&=\mbox{Tr}(\hat{O}\frac{d\rho}{dt})\\
&=i\braket{[H,\hat{O}]}+\mbox{Tr}(\hat{O}\mathcal{D}[L](\rho)),\\
&=i\braket{[H,\hat{O}]}+\braket{\tilde{\mathcal{D}}[L](\hat{O})},
\end{split}
\end{equation}
where $\tilde{\mathcal{D}}[L](\hat{O})$ is called the `adjoint operator', having the following form.
\begin{equation}
\begin{split}
\tilde{\mathcal{D}}[L](\hat{O})&=L^\dag \hat{O}L-\frac{1}{2}\{L^\dag L,\hat{O}\},\\
&=\frac{1}{2}\left(L^\dag[\hat{O},L]+[L^\dag,\hat{O}]L\right).
\end{split}
\end{equation}
Following Eq.~\eqref{odot} and using the mater equation Eq.~\eqref{single_master}, we evaluate the dynamical equation of expectation value of the annihilation operator ($\braket{a}$) as
\begin{equation}
\begin{split}
\dot{\braket{a}}&=i\braket{[\omega a^\dag a,a]}+k_1\braket{\tilde{\mathcal{D}}[a^\dag](a)}+k_2\braket{\tilde{\mathcal{D}}[a^2](a)},\\
&=i\omega \braket{[a^\dag a,a]}+\frac{k_1}{2}\braket{(a[a,a^\dag]+[a,a]a^\dag)}\\
&+\frac{k_2}{2}\braket{({a^\dag}^2[a,a^2]+[{a^\dag}^2,a]a^2)},\\
&=-i\omega\braket{a}+\frac{k_1}{2}\braket{a}-k_2\braket{a^\dag a^2},
\end{split}
\end{equation}
which is similar to Eq.~\eqref{single_amp} as $\braket{a}\equiv\alpha$ and $\braket{a^\dag a^2} \approx |\braket{a^2}|\braket{a} $\cite{lee_pre_1}.

\begin{acknowledgments}
B.B. and T.K. acknowledge the University Grants Commission (UGC), India for providing Junior Research Fellowship. T. B. acknowledges the financial support from the Science and Engineering Research Board (SERB), Govt. of India, in the form of a Core Research Grant [CRG/2019/002632]. 
\end{acknowledgments}

%\bibliography{ref}
\providecommand{\noopsort}[1]{}\providecommand{\singleletter}[1]{#1}%
\end{document}